\documentclass[seceq]{ptptex}

\usepackage{overcite}
\usepackage{graphicx,color}
\title{%        %You can use \\ for explicit line-break
Evolutionary Conditions in the Dissipative MHD System Revisited
}

\author{%       %Use \scshape  for the family name
Tsuyoshi \textsc{Inoue} and Shu-ichiro \textsc{Inutsuka}%
}

\inst{%         %Affiliation, neglected when [addenda] or [errata]
Department of Physics, Graduate Scool of Science, Kyoto University, Sakyo-ku, Kyoto 606-8588, Japan}

\abst{%         %this abstract is neglected when [addenda] or [errata]
The evolutionary conditions for the dissipative continuous magnetohydrodynamic (MHD) shocks are studied.
We modify Hada's approach in the stability analysis of the MHD shock waves.
The matching conditions between perturbed shock structure and asymptotic wave modes shows that all types of the MHD shocks, including the intermediate shocks, are evolutionary and perturbed solutions are uniquely defined.
We also adopt our formalism to the MHD shocks in the system with resistivity without viscosity, which is often used in numerical simulation, and show that all types of shocks that are found in the system satisfy the evolutionary condition and perturbed solutions are uniquely defined.
These results suggest that the intermediate shocks may appear in reality.
}

\begin{document}

\maketitle

\section{Introduction}
The magnetohydrodynamic (MHD) Rankine-Hugoniot relations possess six shock solutions; fast, slow and four types of intermediate shocks, which satisfy entropy condition.
Arguments based upon the theory of strictly hyperbolic systems of conservation laws led to the so-called evolutionary conditions (Lax 1957), and all types of the intermediate shocks are non-evolutionary in the ideal MHD system (Landau \& Lifshitz 1960; Jeffery \& Taniuti 1964; Kantrowiz \& Petschek 1966; Polovin \& Demutskii 1990; and references therein).
These conditions effectively ruled out the intermediate shocks in nature, since they have no neighboring solution corresopnding to small perturbations.

Contrary to this belief, series of numerical experiments in the dissipative MHD system by Wu (1987, 1988, 1990) showed that at least some of the intermediate shocks are admissible and can be formed through nonlinear steeping from continuous waves.
Furthermore, Chao et al. (1993) reported the detection of an interplanetary intermediate shock in the Voyager 1 data.

The dissipative steady solutions that correspond to the fast shock and the slow shock are coplanar (i.e., the velocity field and the magnetic field are in the same plane everywhere).
On the other hand, four types of the dissipative steady solutions of the intermediate shocks can have non-coplanar structure inside their thin but a finite thickness front, and the non-coplanar component magnetic flux is limited by a maximum value, which is proportional to the dissipation coefficients.
These properties were first pointed out by Wu (1990).

The interactions between the intermediate shocks and the Alfv\'en waves were studied by many authors through nonlinear dissipative MHD simulations (Wu 1988; Wu \& Kennel 1992; Markovskii \& Skorokhodov 2000; Falle \& Komissarov 2001).
If classical evolutionary condition is effective even in the dissipative system, the intermediate shocks should instantly disappear.
However, their results do not show such evolution.
Wu (1988) and Wu \& Kennel (1992) showed that the intermediate shock survives a finite time after the interaction with the non-linear Alfv\'en wave whose transverse magnetic field is rotated.
Note that such interaction makes the intermediate shock non-coplanar even in the outsides of the shock structure.
Therefore, they conjectured that there exists a new class of time-dependent intermediate shocks, which do not obey the MHD Rankine-Hugoniot relations, since they violate the coplanarity between the upstream and the downstream, and they are the neighboring states of the intermediate shocks.
Wu (1988) also reported that the intermediate shock remains stable as a result of the interaction with the Alfv\'en wave.
Therefore, the intermediate shocks seem to be evolutionary, but whether the intermediate shock evolves into other shocks and waves or not depends on the nature of the perturbations.
Falle \& Komissarov (2001) showed that the lifetime of the intermediate shock becomes short in a noisy circumstance.
The reason is as follows.
The magnetic flux inside the structure of the intermediate shock is rotated as a result of the interactions with the Alfv\'en waves from the downstream.
Then the intermediate shock eventually brakes up, because of the existence of maximum value of the non-coplanar component magnetic flux that the shock can manage.
Thus, they cautioned that the intermediate shocks tend to survive for a long time in the numerical simulation, because the numerical diffusion makes the maximum value of the non-coplanar magnetic flux large. 

The reanalysis of the evolutionary conditions for the intermediate shocks in the dissipative MHD system are done by Hada (1994) and Markovskii (1998a, b).
Hada (1994) reported that there are additional wave modes, which are originated in the dissipations, and the number of outgoing waves from the shock front is lager than the case of the ideal MHD system.
He concluded that the intermediate shocks are evolutionary, but the set of equations are under-determined due to the excessive dissipative modes.
As a result, he introduced the minimum dissipation principle in order to uniquely define solution.

Previous authors who studied the evolutionary conditions in the dissipative MHD system did not take account the continuous structure of the shock front.
Their analyses are based on the linear perturbation theory of the discontinuities or weak solutions.
However, in the dissipative system, we have to treat the unperturbed shocks as a continuous transition layer, and have to solve differential equations instead of the conservation laws in the ideal system.
In this paper, we examine the evolutionary conditions for the continuous MHD shock waves.
\S 2 provides basic equations that describe perturbation of the shock structure.
In \S 3, we formulate the evolutionary conditions for the continuous shock waves.
We adapt our formulation to the various MHD shocks in \S 4 and \S 5.
In \S 6, we summarize our result and discuss their implication.

\section{Basic Equations for Linear Analysis of Dissipative MHD System}

The one-dimensional, dissipative MHD equations are described as follows:

\begin{equation} \label{basic1}
\frac{\partial \vec{u}}{\partial t}+\frac{\partial \vec{f}}{\partial x}=\frac{\partial \vec{d}}{\partial x},
\end{equation}
\begin{equation}
\vec{u}=\left(
\begin{array}{c}
\rho \\ \rho\,v_{x} \\ \rho\,\vec{v}_{t} \\ \vec{B}_{t} \\ e \\
\end{array} \right),\,
\vec{f}=\left(
\begin{array}{c}
\rho\,v_{x} \\ \rho\,v_{x}^{2}+P-B_{x}^{2} \\ \rho\,v_{x}\,\vec{v}_{t}-B_{x}\,\vec{B}_{t} \\ v_{x}\,\vec{B}_{t}-\vec{v}_{t}\,B_{x} \\ (e+P)\,v_{x}-B_{x}\,(\vec{v}\cdot\vec{B}),
\end{array} \right),
\end{equation}
\begin{equation}
\vec{d}=\left(
\begin{array}{c}
0 \\ \left(\frac{4}{3}\nu+\mu\right)\frac{\partial v_{x}}{\partial x} \\
\nu\,\frac{\partial\vec{v}_{t}}{\partial x} \\
\eta\,\frac{\partial\vec{B}_{t}}{\partial x} \\
\left(\frac{\nu}{3}+\mu\right)v_{x}\frac{\partial v_{x}}{\partial x}+\nu\,\vec{v}\cdot \frac{\partial\vec{v}}{\partial x}+\eta\vec{B}_{t}\cdot \frac{\partial\vec{B}_{t}}{\partial x}+\kappa\frac{\partial}{\partial x}\big( \frac{p}{\rho} \big) \\
\end{array} \right),
\end{equation}
\begin{equation} \label{basic4}
e=\frac{1}{2}\rho\,v^{2}+\frac{p}{\gamma -1}+\frac{1}{2}B^{2},\,\,P=p+\frac{1}{2}B^{2},
\end{equation}
where the subscript $x$ denote the $x$-component and the subscript $t$ denote the $y$ and $z$-component.
We use the unit such that the factor $4\pi$ does not appear.
$\nu$ and $\mu$ are the shear and bulk viscosity coefficients, $\eta$ is the electric resistivity, and $\kappa$ is the heat conduction coefficient.
In addition to (\ref{basic1}), we impose $\boldsymbol{\nabla}\cdot \boldsymbol{B}=0$.
This leads $B_{x}$ to a constant in the one-dimensional case.

In the following, we consider the steady shock solution of equations (\ref{basic1})-(\ref{basic4}) as an unperturbed state.
In this case, without loss of generality, we can choose the coordinate system such that the upstream is $x=-\infty$, the downstream is $x=+\infty$ (i.e., $v_{x,0}>0$ everywhere), and the unperturbed velocity and magnetic fields are in the $x-y$ plane at $x=\pm \infty$ (coplanarity).
We also choose the shock rest frame, and assume that the shock structure is around $x=0$.
We call the steady state type 1 if $v_{x}\ge c_{f}$; type 2 if $c_{f}\ge v_{x}\ge c_{i}$; type 3 if $c_{i}\ge v_{x}\ge c_{s}$; and type 4 if $c_{s}\ge v_{x}$, where $c_{f},\,c_{s}$ and $c_{i}$ are the fast, slow and Alfv\'en (intermediate) speed, respectively.
Steady shock solutions of equations (\ref{basic1})-(\ref{basic4}) are well studied (see, e.g. Wu 1990).
There are six types of the shock solutions, $1\rightarrow 2$ fast shock, $3\rightarrow 4$ slow shock and $1\rightarrow 3$, $1\rightarrow 4$, $2\rightarrow 3$ and $2\rightarrow 4$ intermediate shocks, where the numbers before and after the arrow represent the state of ahead and behind the shock front.

Let us consider the small perturbation of the steady shock structure.
We assume that the perturbation of the physical variable $g(x,t)$ takes the following form:
\begin{equation}
g(x,t)=g_{0}(x)+\delta g(x)\,e^{-i\,\omega t}\,,
\end{equation}
where the subscript $0$ denotes the unperturbed variable.
Linearizing the equation (\ref{basic1}), we obtain the perturbed shock equations
\begin{eqnarray}
\frac{d}{dx}\delta\rho&=&\frac{1}{v_{x,0}}\big( i\,\omega\,\delta\rho-\frac{dv_{x,0}}{dx}\delta\rho-\rho_{0}\frac{d\delta v_{x}}{dx}-\frac{d\rho_{0}}{dx}\delta v_{x} \big), \label{EOC}\\
\Big( \frac{4}{3}\nu+\mu \Big) \frac{d^{2}}{dx^{2}}\delta v_{x} &=& \frac{d}{dx}\big( v_{x,0}^{2}\,\delta\rho+2\,\rho_{0}\,v_{x,0}\,\delta v_{x}+\delta p+B_{y,0}\,\delta B_{y}+B_{z,0}\,\delta B_{z} \big)
\nonumber\\&& -i\,\omega\left(\rho_{0}\,\delta v_{x}+v_{x,0}\,\delta \rho \right), \label{EOMx}\\
\nu\,\frac{d^{2}}{dx^{2}}\delta v_{y} &=& \frac{d}{dx}\big( v_{x,0}\,v_{y,0}\,\delta\rho+\rho_{0}\,v_{y,0}\,\delta v_{x}+\rho_{0}\,v_{x,0}\,\delta v_{y}-B_{x,0}\,\delta B_{y} \big)
\nonumber\\&& -i\,\omega\left(\rho_{0}\,\delta v_{y}+v_{y,0}\,\delta \rho \right), \label{EOMy}\\
\nu\,\frac{d^{2}}{dx^{2}}\delta v_{z} &=& \frac{d}{dx}\big( v_{x,0}\,v_{z,0}\,\delta\rho+\rho_{0}\,v_{z,0}\,\delta v_{x}+\rho_{0}\,v_{x,0}\,\delta v_{z}-B_{x,0}\,\delta B_{z} \big)
\nonumber\\&& -i\,\omega\left(\rho_{0}\,\delta v_{z}+v_{z,0}\,\delta \rho \right), \label{EOMz}\\
\eta\,\frac{d^{2}}{dx^{2}}\delta B_{y} &=& \frac{d}{dx}\big( B_{y,0}\,\delta v_{x}+v_{x,0}\,\delta B_{y}-B_{x,0}\,\delta v_{y} \big)-i\,\omega\,\delta B_{y}, \label{IEy}\\
\eta\,\frac{d^{2}}{dx^{2}}\delta B_{z} &=& \frac{d}{dx}\big( B_{z,0}\,\delta v_{x}+v_{x,0}\,\delta B_{z}-B_{x,0}\,\delta v_{z} \big)-i\,\omega\,\delta B_{z}, \label{IEz}\\
\frac{\kappa}{\rho_{0}} \frac{d^{2}}{dx^{2}}\delta p &=& \kappa\,\Big( \frac{2}{\rho_{0}^{2}}\,\frac{d\rho_{0}}{dx}\,\frac{d\delta p}{dx}-\frac{d^{2}\rho_{0}}{dx^{2}}\,\frac{\delta p}{\rho_{0}^{2}} \Big)+
\kappa\,\frac{d^{2}}{dx^{2}}\Big( \frac{p_{0}}{\rho_{0}^{2}}\,\delta \rho \Big)
\nonumber\\&&-\Big( \frac{4}{3}\nu+\mu \Big) \frac{d}{dx}\Big( v_{x,0}\,\frac{d\delta v_{x}}{dx}+\frac{dv_{x,0}}{dx}\,\delta v_{x} \Big)
\nonumber\\&&-\nu\,\frac{d}{dx}\,\Big( v_{y,0}\frac{d\delta v_{y}}{dx}+\frac{dv_{y,0}}{dx}\,\delta v_{y}+v_{z,0}\,\frac{d\delta v_{z}}{dx}+\frac{dv_{z,0}}{dx}\,\delta v_{z} \Big)
\nonumber\\&&-\eta\frac{d}{dx}\Big( B_{y,0}\frac{d\delta B_{y}}{dx}+\frac{dB_{y,0}}{dx}\delta B_{y}+B_{z,0}\frac{d\delta B_{z}}{dx}+\frac{dB_{z,0}}{dx}\delta B_{z} \Big)
\nonumber\\&&+\frac{d}{dx}\Big\{ \big( \delta e+\delta p+B_{y,0}\delta B_{y}+B_{z,0}\delta B_{z} \big)v_{x,0}+\big( e_{0}+p_{0}
\nonumber\\&&+\frac{1}{2}B_{0}^{2} \big)\delta v_{x} \nonumber-\big( \delta v_{x}\,B_{x,0}+\delta v_{y}\,B_{y,0}+\delta v_{z}\,B_{z,0}+v_{y}\,\delta B_{y,0}
\nonumber\\&&+v_{z}\,\delta B_{z,0}+ \big)B_{x} \Big\}-i\,\omega\,\delta e \,.\label{EE}
\end{eqnarray}

Equation (\ref{EOC}) is the first-order ordinary differential equation with respect to $\delta \rho$, and equations (\ref{EOMx})-(\ref{EE}) are the second-order differential equation with respect to $\delta v_{x},\,\delta v_{y},\,\delta v_{z},\,\delta B_{y},\,\delta B_{z},\,$ and $\delta p$ respectively.
Therefore, we have totally the 13th order differential equations.
We denote the order of the perturbed equations as $N=13$.
If we choose the coplanar intermediate shock solution ($v_{z,0}=B_{z,0}=0$ everywhere) as an unperturbed state, equations (\ref{EOC})-(\ref{EE}) are separated into the two sets of equations.
The first set is equations (\ref{EOC})-(\ref{EOMy}),(\ref{IEy}), and (\ref{EE}), which relate the $x$ and $y$- components of the perturbations, and they are totally the $9$th order differential equations ($N_{xy}=9$).
The other set is equations (\ref{EOMz}) and (\ref{IEz}), which relate the $z$-components of the perturbations, and they are totally the $4$th order differential equations ($N_{z}=4$).

\section{Evolutionary Conditions for the Continuous Shock Waves}
\begin{center}
\begin{figure}
\includegraphics[width=10cm]{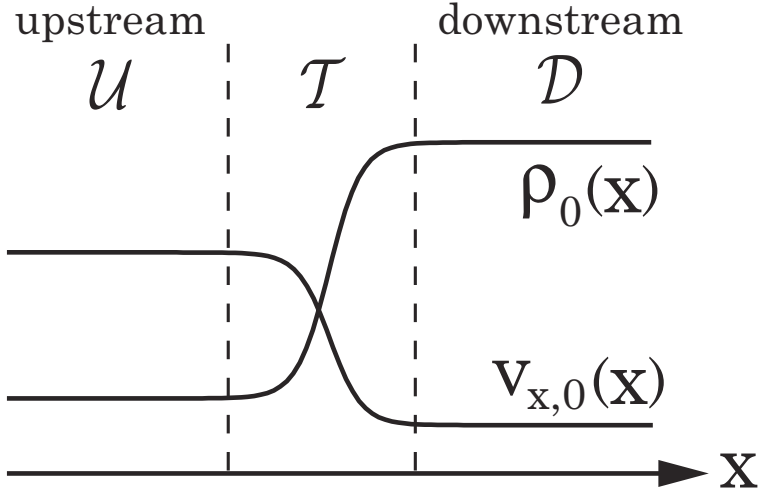}
\caption{Schematic of region partition.
}
\label{fig1}
\end{figure}
\end{center}

In this section, we formulate the evolutionary conditions for the continuous MHD shock waves.
We divide the space into three regions.
Two of them are the regions far away from the shock front, in which the unperturbed physical variables can be regarded as constants.
We call such regions in the upstream and the downstream as region $\cal{U}$ and $\cal{D}$, respectively.
The region between the regions $\cal{U}$ and $\cal{D}$ is the transition region where the unperturbed physical variables change continuously.
We call it region $\cal{T}$.
The situation is schematically illustrated in Figure \ref{fig1}.

As shown in the previous section, perturbed shock equations are totally $N$th order ordinary differential equations.
Thus, in order to determine a solution of these equations in the region $\cal{T}$, we need $N$ boundary conditions at the edges of the region $\cal{T}$.
In other words, in the region $\cal{T}$, there are $N$ degree of freedom in order to determine a solution of these equations.

In the regions $\cal{U}$ and $\cal{D}$, unperturbed physical variables are regarded as constants, and we can obtain the asymptotic solutions of the perturbed shock equations.
Omitting the spatial derivative of the zeroth order variable, and Fourier transforming equations (\ref{EOC})-(\ref{EE}) in space ($\partial/\partial x\sim i\,k$), one can obtain the characteristic equation for the asymptotic waves.
The determinant of the characteristic matrix provides the characteristic equation.
\begin{eqnarray}
D_{fse}\,D_{i}&=&0 \,, \label{disp} \\
D_{fse}&=&c_{i}^{2}( \gamma\,\Omega\,K\,V' + 3\,i\,c_{a}^{2}K' ) \nonumber \\&&+
3\,V\,\{ i\,c_{a}^{2}\,K'\,R + \gamma\,\Omega\,K\,(c_{A}^{2}-c_{i}^{2}+R\,V'/3) \} \,, \label{dispfse} \\
D_{i}&=&\left( \Omega+i\,\eta\,k^{2} \right)
\left( \Omega+i\,\nu\,k^{2} \right) -c_{i}^{2}\,k^{2}\,, \label{dispi}
\end{eqnarray}
where 
\begin{eqnarray}
\Omega &=& \omega - k\,v_{x,0}\,,\\
c_{a}^{2} &=& \gamma\,p_{0}/\rho_{0}\,,\\
c_{i}^{2} &=&  B_{x,0}^{2}/\rho_{0}\,,\\
c_{A}^{2} &=& (B_{x,0}^{2}+B_{y,0}^{2})/\rho_{0}\,,\\
V &=& k^{2}\nu/\rho-i\,\Omega\,,\\
V' &=& k^{2}(3\,\mu-4\,\nu)/\rho-3\,i\,\Omega\,,\\
R &=& \eta-i\,\Omega/k^{2}\,,\\
K &=& (\gamma-1)\,\kappa/\rho-i\,\Omega/k^{2}\,,\\
K'&=& k^{2}(\gamma-1)\,\kappa/\rho-i\,\gamma\,\Omega\,,
\end{eqnarray}
where the zoroth order variables are evaluated in the region $\cal{U}$ and $\cal{D}$, and we use the fact that the non-coplanar component of the unperturbed variables vanish ($B_{z,0}=v_{z,0}=0$) in the asymptotic regions owing to the coplanarity.
The solutions of the characteristic equation $D_{fse}=0$ for $\omega$ as a function of $k$ provide the dispersion relations of the fast, slow, and entropy waves in the uniform dissipative medium, and the solutions of $D_{i}=0$ provide the dispersion relation of the Alfv\'en waves.

\textbf{Definition of Mode }
As shown below, we use the asymptotic solutions in regions $\cal{U}$ and $\cal{D}$ as a boundary conditions of the differential equations (\ref{EOC})-(\ref{EE}).
These conditions are expressed as the superpositions of the independent asymptotic solutions.
The degree of freedom of the spatial behavior of the perturbation for given $\omega$ gives the number of independent asymptotic solutions.
Thus, they are obtained by solving characteristic equation (\ref{disp}) for $k$ as a function of given $\omega$.
In this paper, we call it ``mode" or ``asymptotic mode".
Note that it does not mean solution of characteristic equation for $\omega$ as a function of $k$.

Hada (1994) studied the solutions of $D_{i}=0$ in the limit of small dissipation coefficients.
He found that in addition to the solutions, which correspond to the Alfv\'en modes, there are additional dissipative modes.
Details of the solutions are discussed in the next section.

Let us consider evolutionary condition.
If we steadily throw a small amplitude incident (ingoing) wave whose frequency is $\omega$ toward the shock from the region $\cal{U}$ or $\cal{D}$, then the shock front is perturbed and emit the waves with the same frequency.
Let $m$ $(=m_{\cal{U}}+m_{\cal{D}})$ denote the number of resulting asymptotic modes in the region $\cal{U}$ $(m_{\cal{U}})$ and $\cal{D}$ $(m_{\cal{D}})$ without the incident wave, i.e. the number of modes that can be emitted or raised at the shock front.
In the regions $\cal{U}$ and $\cal{D}$, the solutions of perturbed shock equations should be expressed as the superpositions of the $m$ asymptotic modes and one incident wave.
These asymptotic solutions are determined by $m$ parameters, i.e. amplitudes of the $m$ asymptotic modes.
Note that the amplitude of the incident wave is determined.
Thus, in the case that
\begin{equation}
N=m\,, \label{ECD}
\end{equation}
we can obtain the solution of the perturbed shock equations (\ref{EOC})-(\ref{EE}) in the region $\cal{T}$, which is smoothly connected to the asymptotic solutions at the edges of the region $\cal{T}$, for arbitrary incident wave by choosing $m$ amplitudes of the asymptotic modes.

If $m$ is less than $N$, we cannot obtain solution.
If $m$ is greater than $N$, we can obtain solution, but we need additional conditions or constraints in order to uniquely define the solution.
Therefore, the condition (\ref{ECD}) corresponds to the evolutionary condition in the dissipative system.

If unperturbed shock structure is coplanar, the evolutionary condition (\ref{ECD}) can be divided into the following two conditions
\begin{equation}
N_{xy} = m_{xy}\,, \label{FSECD}
\end{equation}
\begin{equation}
N_{z} = m_{z}\,, \label{ICD}
\end{equation}
where $m_{xy}$ and $m_{z}$ are the numbers of the asymptotic modes of $D_{fse}=0$ and $D_{i}=0$, respectively.
We call the condition (\ref{FSECD}) evolutionary condition for the $x$ and $y$-components, and condition (\ref{ICD}) evolutionary condition for the $z$-components.
Of course, evolutionary shock must satisfy both conditions.

\section{Evolutionary Condition for the $z$-components}

In this section we show that, contrary to the ideal system, all types of intermediate shocks satisfy evolutionary condition for the $z$-components $N_{z}=m_{z}$.
The solutions of $D_{i}=0$ for $k$ under the assumption of small dissipation coefficients, derived by Hada (1994), are
\begin{eqnarray}
k^{(\pm)}&=&\frac{\omega}{v_{x,0}\pm c_{i}}\,+\,\mathcal{O}(\eta,\,\nu)\,, \label{Alfpm} \\
k^{(\pm d)}&=&\frac{-i}{2\,\eta\,\nu}\,\Big( (\eta+\nu)\,v_{x,0}\pm \Big\{(\eta+\nu)^{2}v_{x,0}^{2}-4\,\eta\,\nu\,(v_{x,0}^{2}-c_{i}^{2})\Big\}^{\frac{1}{2}}\Big)\,+\,\mathcal{O}(\eta^{0},\,\nu^{0})\,. \label{Dpm}
\end{eqnarray}
The $(+)$ and $(-)$ modes are the usual Alfv\'en waves, which propagate parallel and anti-parallel to the $x$-axis in the fluid rest frame.
The $(\pm d)$ modes, with no counterparts in the ideal system, are so called the ``dissipative modes".
%It is worth notice again that the ``mode" used in this paper means the mode of the spatial behavior of the perturbation for given $\omega$, i.e. the solution of the dispersion relation for $k$.
The dissipative modes do not propagate, because $k^{(\pm d)}$ do not have real part in their primary terms, and they are localized around the shock front.
We exclude diverging dissipative modes as unphysical asymptotic solutions.
Thus, they are physical if
\begin{equation}
\mbox{Im}[\,k^{(\pm d)}\,]>0 \,\mbox{ in the region } \cal{U}, \mbox{ or}
\end{equation}
\begin{equation}
\mbox{Im}[\,k^{(\pm d)}\,]<0 \,\mbox{ in the region } \cal{D}.
\end{equation}
If the flow speed is super-Alfv\'enic in the shock rest frame, then from equation (\ref{Dpm}), the imaginary part of $k^{(\pm d)}$ are less than zero.
Therefore there are two dissipative modes in the region $\cal{U}$, whereas no dissipative mode in the region $\cal{D}$. On the other hand, if the flow speed is sub-Alfv\'enic, the imaginary part of $k^{(+d)}$ is greater than zero and that of $k^{(-d)}$ is less than zero, thus there is one dissipative mode in the region $\cal{U}$ and $\cal{D}$.

\begin{center}
\begin{figure}
\includegraphics[width=10cm]{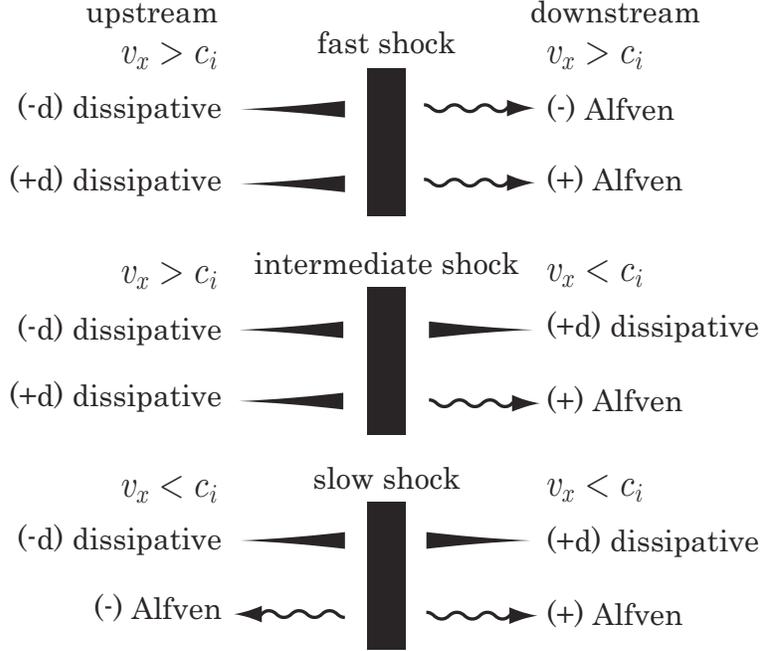}
\caption{Illustration of the outgoing ordinary MHD modes and the localized dissipative modes, where the bars represent the region $\cal{T}$, the arrows represent the propagation of the asymptotic modes, and the triangles represent the localized dissipative modes.
The ($\pm$) sign before the Alfv\'en modes denote the propagation direction in the fluid rest frame, and
$c_{i}=(B_{x,0}^{2}/\rho_{0})^{1/2}$ is the Alfv\'en (intermediate) speed.
}
\label{fig2}
\end{figure}
\end{center}

Let us consider the evolutionary condition for the $z$-components.
In the case of the fast shock, the flow speeds both in the upstream and the downstream are super-Alfv\'enic. 
Therefore, the $k^{(\pm)}$ modes propagate to the downstream in the region $\cal{D}$, and the $k^{(\pm d)}$ modes are in the region $\cal{U}$.
In the same way, the flow speeds of the slow shock are sub-Alfv\'enic in the both sides. There are $k^{(-)}$ and $k^{(+d)}$ modes in the region $\cal{U}$, and $k^{(+)}$ and $k^{(-d)}$ modes in the region $\cal{D}$.
In the case of the intermediate shocks, the upstream flow speed is super-Alfv\'enic and the downstream flow speed is sub-Alfv\'enic.
There are $k^{(\pm d)}$ modes in the region $\cal{U}$, and $k^{(+)}$ and $k^{(-d)}$ modes in the region $\cal{D}$. 
The situations are schematically illustrated in Figure \ref{fig2}.
In all cases, the MHD shocks satisfy the evolutionary condition for the $z$-components, i.e. $N_{z}=m_{z}=4$.

\section{Evolutionary Conditions in the Resistive but Inviscid MHD System}

Let us consider the case where there is only resistivity in the dissipation.
The one-dimensional basic equation of the weakly ionized gas, a main ingredient of the interstellar medium, can be written as a resistive but inviscid MHD equation under the strong coupling (one-fluid) approximation.
Furthermore, Wu (1987, 1990) showed by using one-dimensional simulation in this system that the intermediate shocks are formed through nonlinear steeping from simple waves.
Thus, it is meaningful to analyze the evolutionary condition in this system.
We apply our formalism developed in \S 3 to the resistive but inviscid MHD shocks.

In this system the perturbed basic equations (\ref{EOC})-(\ref{EE}) become first order ordinary differential equations with respect to $\delta \rho,\,\delta p,\,\vec{\delta v}$, and second order one with respect to $\vec{\delta B}_{t}$.
Thus, we have totally 9-th order ($N=9$) differential equations $(N_{xy}=6,\,N_{z}=3$ in the case of the coplanar shocks$)$.
The number of asymptotic modes is determined by the solutions of the resistive version of equation (\ref{disp}).
\begin{eqnarray}
D_{fse}&=&\Omega\,\{\,c_{i}^{2}\,c_{a}^{2}\,k^{4}-i\,\eta\,c_{a}^{2}\,\Omega\,k^{4}-(c_{A}^{2}+c_{a}^{2})\Omega^{2}\,k^{2}+i\,\eta\,\Omega^{3}k^{2}+\Omega^{4}\,\} \label{disprfse}\\
D_{i}&=&\Omega\left( \Omega+i\,\eta\,k^{2} \right)-c_{i,0}^{2}\,k^{2}\,. \label{dispri}
\end{eqnarray}
Equation (\ref{dispri}), which contains three asymptotic modes, was studied by Hada (1994).
Two of them corresponds to two Alfv\'en waves, and third one corresponds to the dissipative mode.
The dissipative mode solution under the assumption of small resistivity is
\begin{equation}
k^{(d1)}=i\,\frac{c_{i}^{2}-v_{x,0}^{2}}{\eta\,v_{x,0}}\,+\,\mathcal{O}(\eta^{0},\,\nu^{0})\,, \label{dissmi}
\end{equation}
which is physical in the upstream for the super-Alfv\'enic state (state 1 and 2), and physical in the downstream for the sub-Alfv\'enic state (state 3 and 4).

Equation (\ref{disprfse}) contains six asymptotic modes.
Five of which correspond to two fast waves, two slow waves and one entropy wave, and the last one corresponds to the dissipative mode.
The dissipative solution under the assumption of small resistivity is
\begin{equation}
k^{(d2)}=i\,\frac{c_{a}^{2}\,c_{i}^{2}-(c_{A}^{2}+c_{a}^{2})\,v_{x,0}^{2}+v_{x,0}^{4}}{\eta\,v_{x,0}\,(c_{a}^{2}-v_{x,0}^{2})}\,+\,\mathcal{O}(\eta^{0},\,\nu^{0})\,. \label{dissmfse}
\end{equation}
In the case of the MHD system with resistivity without viscosity, steady shock structure whose flow velocity is supersonic ($v_{x}>c_{a}$) ahead of the shock front and subsonic ($v_{x}<c_{a}$) behind the shock front is impossible (see, Wu 1990).
Thus, it is possible that the shock structure for $1\rightarrow2$, $1\rightarrow3$, and $2\rightarrow3$ whose flow speeds are supersonic everywhere in the shock rest frame, and $2\rightarrow3$, $2\rightarrow4$, and $3\rightarrow4$ whose flow speeds are subsonic everywhere in the shock rest frame.
The denominator of the righthand side of equation (\ref{dissmfse}) is less than zero in the supersonic state, and larger than zero in the subsonic state.
The numerator is larger than zero in the state 1 (super-fast) and 4 (sub-slow), and less than zero in the state 2 and 3 (sub-fast and super-slow).
Therefore, in the supersonic case, the dissipative (d2) mode is physical in the region $\cal{U}$ for the state 1, and in the region $\cal{D}$ for the state 2 and 3.
In the subsonic case, it is physical in the $\cal{U}$ for the state 2 and 3, and in the $\cal{D}$ for the state 4.

\begin{figure}
\includegraphics[width=10cm]{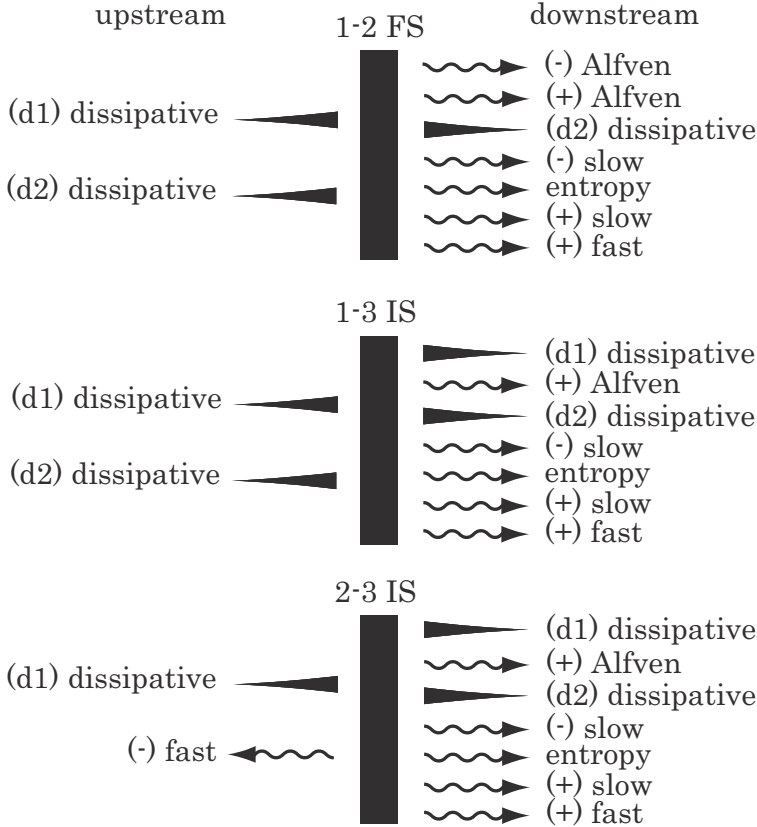}
\caption{
Illustration of the outgoing ordinary MHD modes and the localized dissipative modes for the shocks whose upstream velocity is supersonic, where the bars represent the region $\cal{T}$, the arrows represent the propagation of the asymptotic modes, and the triangles represent the localized dissipative modes.
The ($\pm$) sign before the ordinary MHD modes denote the propagation direction in the fluid rest frame.
}
\label{fig3}
\end{figure}

\begin{figure}
\includegraphics[width=10cm]{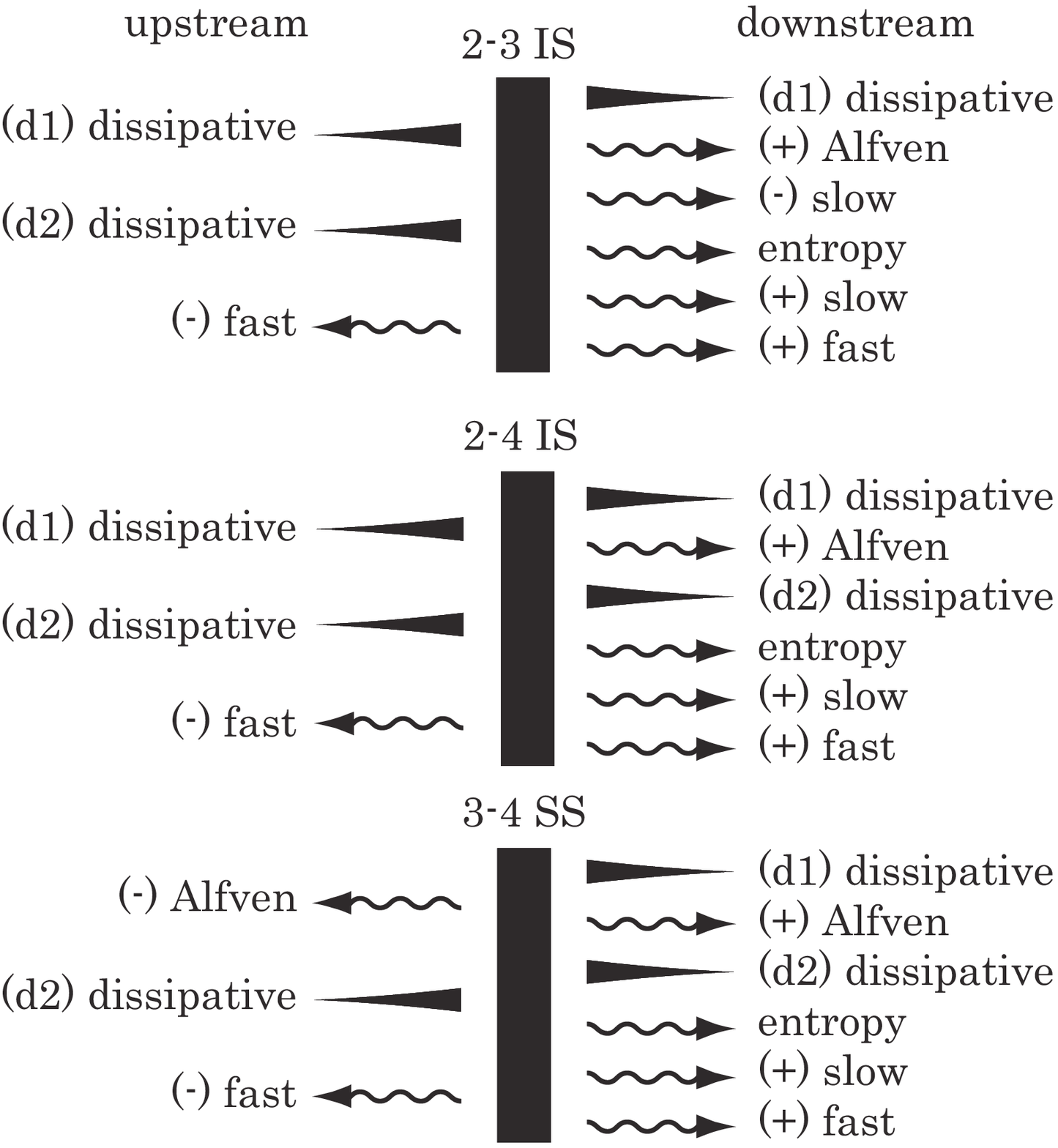}
\caption{
Illustration of the outgoing ordinary MHD modes and the localized dissipative modes for the shocks whose upstream velocity is subsonic, where the bars represent the region $\cal{T}$, the arrows represent the propagation of the asymptotic modes, and the triangles represent the localized dissipative modes.
The ($\pm$) sign before the ordinary MHD modes denote the propagation direction in the fluid rest frame.
}
\label{fig4}
\end{figure}

The number of asymptotic modes in asymptotic regions is $m=9\,\,(m_{xy}=6,\,m_{z}=3$ in the case of non-coplanar shock$)$ for all shocks.
%We illustrate the situations for the supersonic shocks in Figure \ref{fig3}, and for the subsonic shocks in Figure \ref{fig4}.
We illustrate the situations for the shocks whose upstream velocity is supersonic in Figure \ref{fig3}, and for the shocks whose upstream velocity is subsonic in Figure \ref{fig4}.
Therefore, all the shocks, which are allowed in the MHD system with resistivity without viscosity, satisfy the evolutionary condition $N=m=9\,\,($or $N_{xy}=m_{xy}=6,\,N_{z}=m_{z}=3$). 

\section{Summary and Discussion}
We have studied evolutionary conditions for the continuous MHD shock waves in the dissipative system.
We have shown that all types of MHD shocks, even the intermediate shocks, satisfy the evolutionary condition for the $z$-components, which is not satisfied in the ideal MHD system.
Especially in the resistive system, all types of MHD shocks, which are allowed in the resistive system, satisfy the evolutionary condition.
Therefore, the intermediate shocks have neighboring solution, and they can survive interactions with arbitrary small amplitude waves.

The difference of the evolutionary conditions between the ideal and the dissipative system is that, in the dissipative system, there are dissipative perturbation modes, which do not propagate and are localized around the shock front, and they provide degree of freedom, which are lost in the ideal system.
This difference is first pointed out by Hada (1994).
Hada (1994) tried to connect the perturbation by using the Rankine-Hugoniot relations.
As a result, set of equations become under-determined due to the excessive dissipative modes.
In order to uniquely define solution, Hada (1994) introduced the minimum dissipation principle.
In this paper, however, we have shown that perturbations in the asymptotic regions (far away from the shock front) can be uniquely connected by solving differential equations, instead of the conservation laws in the ideal or hyperbolic system, because the shock is not a weak solution but a continuous solution in the dissipative system.
Note that the solution of the differential equations also satisfies the usual conservation laws in the asymptotic regions.

However, the idea of the minimum dissipation principle and use of the perturbed Rankine-Hugoniot relations may be important, when we analyze the linear stability of the intermediate shocks.
In order to analyze the linear stability by using our formulation, we have to solve the differential equation.
This seems difficult, because the equations are stiff, and there need huge amount of calculations to search parameter space.
If the minimum dissipation principle can sort out important dissipative modes from other dissipative modes in order to the solution be unique, the analysis is easily done, because the solution is obtained analytically.
Further study is needed to discover such technique.

Our approach is similar to that of Wu (1988) and Wu \& Kennel (1992).
They studied the interaction between the intermediate shock and the non-linear Alfv\'en wave whose transverse magnetic field is rotated.
They showed by using non-linear simulations that there exists a new class of time-dependent intermediate shocks, since they violate the coplanarity even without the shock structure (i.e., they do not obey the MHD Rankine-Hugoniot relations between the upstream and the downstream), which possibly be a neighboring state of the intermediate shocks.
On the other hand, our approach treat the interaction between the intermediate shock and small amplitude (linear) MHD wave, and showed that the intermediate shocks have neighboring solutions which describe the time evolutions of the perturbed intermediate shocks in the linear regime.

The intermediate shock which has maximum non-coplanar magnetic flux inside the shock structure break up into other shocks and waves as a result of the interaction with the Alfv\'en wave (Markovskii \& Skorokhodov 2000 and Falle \& Komissarov 2001).
Such an unstable phenomenon would be analyzed possibly in terms of the linear stability analysis.
Our formulation is also expected to be used for such analysis.

The existence of the intermediate shocks has been argued mainly in the interplanetary system (Chao et al. 1993; Chao 1995).
However, in the partially ionized interstellar medium, it is known that there are C-type MHD shocks, which have broad shock structure owing to friction between neutral and ionized gases.
Because of the lifetime of the intermediate shock seems to be proportional to the width of the front, the intermediate shocks can be long-lived in the interstellar medium.
The quantitative discussion on this issue remains to be done.

\section*{Acknowledgements}
This work is supported by the Grant-in-Aid for the 21st Century COE "Center for Diversity and Universality in Physics" from the Ministry of Education, Culture, Sports, Science and Technology (MEXT) of Japan.
SI is supported by the Grant-in-Aid (No.15740118, 16077202, 18540238) from MEXT of Japan.

%\appendix
%\section{First Appendix} %Empty argument \section{} yields `Appendix'. 
%
%\section{Second Appendix}


\begin{thebibliography}{99}
%%%%%%%%%%%%%%%%%%%%%%%%%%%%%%%%%%%%%%%%%%%%%%%%%%%%%%%%%%%%%
% Some macros are available for the bibliography:
%  o for general use
%    \JL : general journals                 \andvol : Vol (Year) Page
%  o for individual journal 
%    \AJ   : Astrophys. J.           \NC         : Nuovo Cim.
%    \ANN  : Ann. of Phys.           \NPA, \NPB  : Nucl. Phys. [A,B]
%    \CMP  : Commun. Math. Phys.     \PLA, \PLB  : Phys. Lett. [A,B]
%    \IJMP : Int. J. Mod. Phys.      \PRA - \PRE : Phys. Rev. [A-E]     
%    \JHEP : J. High Energy Phys.    \PRL        : Phys. Rev. Lett.
%    \JMP  : J. Math. Phys.          \PRP        : Phys. Rep.
%    \JP   : J. of Phys.             \PTP        : Prog. Theor. Phys.     
%    \JPSJ : J. Phys. Soc. Jpn.      \PTPS       : Prog. Theor. Phys. Suppl.
% Usage:
%  \PRD{45,1990,345}          ==> Phys.~Rev.\ \textbf{D45} (1990), 345
%  \JL{Nature,418,2002,123}   ==> Nature \textbf{418} (2002), 123
%  \andvol{B123,1995,1020}    ==> \textbf{B123} (1995), 1020
%%%%%%%%%%%%%%%%%%%%%%%%%%%%%%%%%%%%%%%%%%%%%%%%%%%%%%%%%%%%%
  
\bibitem{C93} J. K. Chao, L. H. Lyu, B. H. Wu, A. J. Lazarus and T. S. Chang, J. Geophys. Res. \textbf{98} (1993), 17433.
\bibitem{C93} J. K. Chao, Adv. Space Res. \textbf{15} (1995), 521.
\bibitem{FK01} S. A. E. G. Falle and S. S. Komissarov, J. Plasma Phys. \textbf{65} (2001), 29.
\bibitem{H94} T. Hada, Geophys. Res. Lett. \textbf{21} (1994), 2275.
\bibitem{JT64} A. Jeffrey and T. Taniuti, \textit{Nonlinear Wave Propagation}, (Academic Press, New York, 1964).
\bibitem{KP66} A. R. Kantrowitz and  H. E. Petschek, MHD characteristics and shock waves, \textit{Plasma Physics in Theory and Application}, ed. W. B. Kunkel, pp. 148, (McGraw-Hill, New York, 1966).
\bibitem{LL} L. D. Landau, and E. M. Lifshitz, \textit{Electrodynamics of continuous media}, (Addison-Wesley, Reading, Mass, 1960).
\bibitem{L57} P. D. Lax, Commun. Pure Appl. Math. \textbf{10} (1957), 537.
\bibitem{M1998a} S. A. Markovskii, Phys. Plasma. \textbf{5} (1988a) 2596.
\bibitem{M1998b} S. A. Markovskii, J. Exp. Theor. Phys. \textbf{86} (1988b) 340.
\bibitem{MS2000} S. A. Markovskii and S. L. Skorokhodov, Phys. Plasmas, \textbf{7} (2000) 158.
\bibitem{PD} R. V. Polovin and V. D. Demutskii, \textit{Fundamentals of Magnetohydrodynamics}, (Consultants Bureau, New York, 1990)
\bibitem{W87} C. C. Wu, Geophys. Res. Lett. \textbf{14} (1987), 668.
\bibitem{W88} C. C. Wu, J. Geophys. Res. \textbf{93} (1988), 987.
\bibitem{W90} C. C. Wu, J. Geophys. Res. \textbf{95} (1990), 8149.
\bibitem{WK92} C. C. Wu and C. F. Kennel, Geophys. Res. Lett. \textbf{19} (1992), 2087.
\end{thebibliography}
\end{document}